%Paper: hep-th/9510129
%From: Cobi Sonnenschein <COBI@taunivm.tau.ac.il>
%Date: Wed, 18 Oct 95 13:13:15 IST

\input harvmac
\parindent0pt
\def\hb{\hfill\break}
\def\p{\partial}
\def\F{{\cal F}}
\def\L{\Lambda}
\def\A{{\cal A}}
\def\a{\alpha}
\def\b{\beta}
\def\l{\lambda}

%file: dfdu.tex\hfill 20.8.95
%\vskip.5cm

%%%%%%%%%%%%%%%%%%% %%%%%%%%%%%%%%%%%%%%%%%%%%%%%%%%%%%%%%%%%%%%%%
%%%%% References
%%%%%%%%%%%%%%%%%%% %%%%%%%%%%%%%%%%%%%%%%%%%%%%%%%%%%%%%%%%%%%%%%
\newif\ifnref
\def\rrr#1#2{\relax\ifnref\nref#1{#2}\else\ref#1{#2}\fi}
\def\ldf#1#2{\begingroup\obeylines
\gdef#1{\rrr{#1}{#2}}\endgroup\unskip}

\nreffalse

\ldf\SW{N. Seiberg and E. Witten, Nucl. Phys. B426 (1994) 19
and B431 (1994) 484}
\ldf\puresun{A. Klemm, W. Lerche, S. Theisen and S. Yankielowicz,
Phys. Lett. B344 (1995) 169 and hep-th/9412158;\hb
C.P. Argyres and A. Farragi, Phys. Rev. Lett. 74 (1995) 3931;\hb
A. Klemm, W. Lerche and S. Theisen, hep-th/9505150}
\ldf\mattersun{P.C. Argyres, R. Plesser  and A. Shapere,
hep-th/9505100;\hb
J. Minahan and D. Nemeshansky, hep-th/9507032}
\ldf\pureson{U. Danielson and B. Sundborg, hep-th/9504102;\hb
A. Brandhuber and K. Landsteiner, Phys. Lett.B 358 (1995) 73}
\ldf\matterson{A. Hanany, hep-th/9509176}
\ldf\matterall{P.C. Argyres and A. Shapere, hep-th/9509175}
\ldf\rs{See e.g. P. Griffiths and J. Harris,
Principles of Algebraic Geometry, John Wiley \& Sons, 1978, Chap. 2.2;\hb
G. Springer, Introduction to Riemann Surfaces,
Addison-Wesley, 1957, Chap. 10.3}
\ldf\PF{A. Klemm, W. Lerche and S. Theisen, hep-th/9505150;\hb
K. Ito and S.-K. Yang, hep-th/9507144}
\ldf\KKLMV{S. Kachru, A. Klemm, W. Lerche, P. Mayr and C. Vafa,
hep-th/9508155}
\ldf\Matone{M. Matone, hep-th/9506102 and hep-th/9506181}
\ldf\sv{M. Shifman and A. Vainshtein, Nucl. Phys. B359 (1991) 571}
\ldf\nsvz{V. Novikov, M. Shifman, A. Vainshtein and V. Zakharov,
Nucl. Phys. B229 (1983) 381 and 407;\hb
G.C. Rossi and G. Veneziano, Phys. Lett. 138B (1984) 195}
\ldf\IS{K. Intriligator and N. Seiberg, B431 (1994) 551 and Nucl. Phys.
and B444 (1995) 125.}
\ldf\EFGR{S. Elitzur, A. Forge, A. Giveon and E. Rabinovici,
hep-th/9504080,
Phys.Lett. B353 (1995) 79 and hep-th/9509130}
\ldf\Hoz{A. Hanany and Y. Oz, hep-th/9505075}

%\footline{\hss\tenrm--\folio\--\hss}
%%%%%%%%%%%%%
\Title{hep-th/9510129, TAUP-2296-95, LMU-TPW-95-15}
{\vbox{\centerline{On the Relation Between the Holomorphic
Prepotential}
\centerline{{\it and}}
\centerline{ the Quantum Moduli in SUSY Gauge
Theories }}\footnote{$^\#$} {Work supported in part by
GIF - the German-Israeli Foundation for Scientific Research}}
\centerline{ J. Sonnenschein$^{*}$,
S. Theisen$^{\dagger}$ and S. Yankielowicz$^{*}$}
\vglue .5cm
\centerline{$^{*}$ School of Physics and Astronomy}
\centerline{Beverly and Raymond Sackler Faculty of Exact Sciences}
\centerline{Tel--Aviv University}
\centerline{Ramat--Aviv, Tel--Aviv 69978, Israel}
\vglue .5cm
\vglue .5cm
\centerline{$^{\dagger}$Sektion Physik}
\centerline{Universit\"at M\"unchen}
\centerline{Theresienstra\ss e 37, 80333 M\"unchen, FRG}
\vglue .5cm
\vglue .5cm

\noindent
We give a simple proof of the relation
$\L\p_\L\F={i\over2\pi}b_1\langle\Tr\phi^2\rangle$,
which is valid for
$N=2$ supersymmetric QCD with massless quarks.
We consider $SU(N_c)$ gauge theories as well as $SO(N_c)$ and $SP(N_c)$.
Aa analogous  relation which corresponds to massive hypermultiplets
is written down.
We also discuss the generalizations to  $N=1$ models in the Coulomb phase.

\vfill\eject

\footline{\hss\tenrm--\folio\--\hss}

A lot of activity has followed the beautiful work of
Seiberg and Witten \SW\
on the exact non-perturbative low energy effective action
(in the Coulomb phase) of the pure and QCD-like $SU(2)$ $N=2$
supersymmetric gauge theories. In \puresun\ it was generalized to
$SU(N_C)$ $N=2$ theories and in \mattersun\Hoz\ to
$SU(N_C)$ $N=2$ theories with matter in the fundamental representation.
Recently this work has been extended to $SO(N_C)$ and $Sp(N_C)$ gauge
groups\pureson\matterson\matterall.

In the present letter we prove and discuss relations between
the prepotential
${\cal F}$ and the quantum moduli of the $N=2$ theory.
The most interesting
relation reads
\eqn\aa{
\L{\p\over\p\L}{\cal F}={i\over2\pi}b_1 \langle\Tr\phi^2\rangle
}
where $\phi$ is the adjoint complex
scalar in the $N=2$ gauge multiplet, and $b_1$ is the one-loop
coefficient of the beta-function. This relation holds for all
$N=2$ theories,
either pure or with massless matter quarks. For the case of
pure $SU(2)$ this
relation is essentially proven in \Matone\ where the modular
transformations of the prepotential ${\cal F}$ are considered. In
\KKLMV\ the generalization of the Seiberg-Witten approach to $N=2$ string
theory is investigated. In particular, the exact non-perturbative
result on pure
$SU(2)$ and $SU(3)$ $N=2$ Yang-Mills theory were recovered
from the tree-level
Type II string theory at the corresponding points in moduli space,
in the limit
of $\a'\to 0$, where gravity is decoupled. In this work
it was observed that
starting from the local case $u\equiv{1\over2}\langle\Tr\phi^2\rangle$
behaves as a period and the relation \aa\
holds with the dilaton playing the role of $\L$. This relation
turns out to be
crucial in obtaining the rigid theory from the local one.

In the pure $N=2$ gauge theory,
the low energy effective action up to terms with two
derivatives is completely determined by one holomorphic function of
$N=2$ chiral superfields $\A_i$, the prepotential ${\cal F}(\A)$.
For $N_f>0$, we also have to include (matter) hypermultiplets, whose
contribution to the low energy effective action
is not determined by a holomorphic structure.
However, for the purpose of this note, we won't need their couplings.
For the massless case, the perturbative piece of the prepotential is
\eqn\fpert{
%\F_{\rm pert.}(\A)=-{i\over8\pi}{2N_C-N_f\over N_C}
%\sum_{\a>0}(\vec\a\cdot\vec\A)^2\ln
\Bigl({(\vec\a\cdot\vec\A)^2\over\L^2}\Bigr)
\F_{\rm pert.}(\A)={1\over2\pi i}{l({\rm adj.})-\sum_i l_i({\rm matter})
\over l({\rm adj.})}
\sum_{\a>0}(\vec\a\cdot\vec\A)^2
\ln\Bigl({(\vec\a\cdot\vec\A)^2\over\L^2}\Bigr)
}
The sum is over all positive roots and $l({\rm adj.})$ is the
index of the adjoint representation of the gauge group $G$
whereas $l_i({\rm matter})$ is the
index of the representation of the $i$'th matter hypermultiplet.
{}From this expression the perturbative beta-function, which is purely
one-loop, follows.

The prepotential may be considered as
a holomorphic function of the chiral superfields $\A_i$ {\it and}
the scale $\Lambda$.
Defining $a_i=\A_i|_{\theta=0}$ and
$a_{D_i}={\p\F(a)\over\p a_i}$, one then finds that $(a_i,a_{D_i})$
are the periods of an abelian differential of the second kind
(having poles with zero residue)
for the case of $N_f\geq0$ {\sl massless} hypermultiplets or of
the third kind (having poles with non-zero residue)
for $N_f>0$ {\sl massive} hypermultiplets.
These differentials are defined on an (auxiliary) hyperelliptic
Riemann surface
$\Sigma_{r}$ of genus $r={\rm rank}(G)$
and the periods are with respect to a symplectic homology basis
with one-cycles $(\a_i,\b_i)$.
The Riemann surfaces for pure $SU(N_C)$ \puresun,
$SU(N_C)$ with hypermultiplets \mattersun\Hoz,
$SO(N_C)$ without \pureson\ and with \matterson\matterall\ matter,
and finally
also for $Sp(N_C)$ \matterall\ have been found by now.
In particular ref.\matterall\ gives curves with genus equal to the rank of $G$.
The hypermultiplets were always chosen in the defining
representation and their number such that the theory is either
asymptotically free or has vanishing beta function.
Recently curves for certain $N=1$ supersymmetric theories were considered
in \IS\Hoz\EFGR\ with matter in the adjoint and/or fundamental representations.
We first treat $N=2$ theories with
$G=SU(N_C)$. The remaining classical groups and some $N=1$ cases
will be dealt with below.

The Riemann surface for $SU(N_C)$ is the genus $N_C-1$ hyperelliptic
curve $\Sigma_{N_C-1}$
\eqn\susurface{
y^2=W^2+F
}
where
\eqn\suW{
W=\langle\det(x{\bf 1}-\phi)\rangle\equiv
x^{N_C}-\sum_{k=2}^{N_C}s_k x^{N_C-k}
}
$F=F(x,m_j,\Lambda)$ is a polynomial of its arguments,
independent of the $s_i$ and
$F(x)\sim x^{N_f}$ for large $x$. If we parametrize $\langle\phi\rangle=
\sum_i a_i H_i$ where $H_i$ are the generators in the Cartan subalgebra,
we get in the semiclassical limit
$s_2={1\over2}a_i a_j\Tr(H_i H_j)$.
The exact (non-perturbative) expression is
$s_2=u={1\over2}\langle\Tr\phi^2\rangle$
where $\phi$ is the Higgs field, i.e. the scalar component of the $N=1$
chiral superfield contained in the $N=2$ chiral superfield.

The meromorphic differential $\lambda$ is
\mattersun\Hoz\foot{Here and below
relations between abelian differentials are always up to exact
differentials}(the prime denotes differentiation w.r.t. $x$)
\eqn\sulambda{
\lambda={1\over2\pi i}(W F'-2 FW'){(x+b)\,dx\over F y}
}
where the normalization is chosen such that ($i=1,\dots, N_C-1$)
$a_i=\int_{\a_i}\l$, $a_{D_i}=\int_{\b_i}\l$ and
$\p_{s_k}a_i=\int_{\a_i}\omega_k$, $\p_{s_k}a_{D_i}=\int_{\b_i}\omega_k$.
$\omega_k={\p\over\p s_k}\lambda=
{1\over \pi i}{x^{N_C-k}dx\over y}$, $k=2,\dots,N_C$,
are a basis of holomorphic differentials (abelian differentials
of the first kind) on $\Sigma_{N_C-1}$.
The constant $b=b(\L,m)$ must be chosen such that for the massless
case there are no poles at zeroes of $F$ and the pole at infinity
has zero residue. In the massive case $\lambda$ must have poles at the zeroes
of $F$ with residues $m_j$.
One finds that in the massless case
$b=0$. $\lambda$ also has a double pole at
infinity with residue $-\sum m_j$ which vanishes in the massless case.
It is, therefore, an abelian differential of the second and third kind
in the massless and massive cases, respectively.

The effective (field dependent, dimensionless) gauge coupling
is given by the matrix $\tau_{ij}={\p^2\F\over\p a_i\p a_j}$.
$\F$ is thus a homogeneous function of weight two of $a_i,m_j,\L$
and satisfies the Euler equation\foot{Here and below, $\L$ is always meant
to be $\L_{N_f}$.}
\eqn\euler{
2\F=(\L\p_\L+\sum_j m_j\p_{m_j}+\sum_i a_i\p_{a_i})\F
}
Taking derivatives w.r.t. to $s_k$ and using the definition of
the $a_{D_i}$
one obtains
\eqn\Euler{
{\p\over \p{s_k}}(\L\p_\L+\sum_j m_j\p_{m_j})\F
=\sum_i(a_i{\p\over\p s_k}a_{D_i}-a_{D_i}{\p\over\p s_k}a_i)
}
Using now the above results we arrive at
\eqn\Riemann{
{\p\over \p{s_k}}(\L\p_\L+\sum_j m_j\p_{m_j})\F
=\sum_i\int_{\a_i}\l\,\int_{\b_i}
\omega_k-\int_{\b_i}\l\,\int_{\a_i}\omega_k
}
The right hand side of this equation can be evaluated with the help of
a Riemann bilinear relation \rs.
Since they make a distinction between $\l$ being abelian of second or third
kind, we will treat the massless and massive cases separately.
We first discuss the massless case, where the integrals on the right hand
side of  eq.\Riemann\ can be done explicitly.
The mass dependent
terms on the left hand side of eqs.\Euler\ and \Riemann\ are
now absent and
$\l$ has a double pole at $w=1/x=0$ with expansion
\eqn\lambdaasymp{
\lambda=(\lambda_{-2}w^{-2}+\lambda_0+\lambda_1 w+\dots)dw
}
with $\lambda_{-2}={1\over2\pi i}(2N_C-N_f)$; there are no further poles
of $\lambda$. The Riemann bilinear relation now reads
\eqn\masslessriemann{
\sum_i\,\int_{\a_i}\lambda\,\int_{\beta_i}\omega_k-\int_{\beta_i}\lambda\,
\int_{\a_i}\omega_k=2\pi i\,
\sum_{n\geq2}{\lambda_{-n}\omega^{(k)}_{n-2}\over n-1}
}
where $\omega^{(k)}_n$ are the coefficients of $\omega_k$ in its expansion
around infinity:
\eqn\omegaasymp{
\omega_k=(\omega^{(k)}_0+\omega^{(k)}_1 w+\dots)dw
=\bigl(-{1\over i\pi}w^{k-2}+O(w^{k-1})\bigr)
}
i.e. $\omega_0^{(k)}=-{1\over \pi i}\delta_{k,2}$.
We then have
\eqn\dfds{
\p_{s_k}(\L\p_\L\F)=2\pi i \lambda_{-2}\, \omega_0^{(k)}
={i\over\pi}(2N_C-N_f)\delta_{k,2}
}
Integration gives
\eqn\dfdl{
\L\p_\L\F={i\over\pi}(2N_C-N_f)s_2
}
where comparison with the weak coupling expression shows that a possible
contribution $const.\,\Lambda^2$ is absent from the right hand side.
Let us briefly comment on this result.
Taking derivatives with respect to $a_i$ and $a_j$ and using the
definition $\p_{a_i}\p_{a_j}{\cal F}=\tau_{ij}
={1\over 2\pi}\theta_{ij}+4\pi i({1\over g^2})_{ij}$ one obtains
\eqn\ldtdl{
\L{d\over d\L}\tau_{ij}={i\over 2\pi}(2N_C-N_f)\p_{a_i}\p_{a_j}
\Tr\langle\phi^2\rangle
\simeq{i\over\pi}(2N_C-N_f)\Tr(H_i H_j)}
where in the last step we have taken the semi-classical limit, i.e. have
suppressed instanton corrections.
\medskip
We note that the relation \ldtdl\ is compatible with perturbation theory.
It is well known \sv\ that ${\cal F}$ (or, equivalently, the
Wilsonian field dependent gauge coupling) acquires a contribution only
at one loop level. This means that $s_2$ is equal (up to nonperturbative
contributions) to its classical value. This agrees with the
general observation that correlators
of lowest components of gauge invariant chiral superfields are
`topological', i.e. they do not depend on positions \nsvz.
Thus they get contributions only from disconnected diagrams.
Moreover, they
depend holomorphically on the parameters, notably on the gauge coupling.
This in fact implies (since there is no dependence on $\theta$ in
perturbation theory) that there are no perturbative quantum corrections
to the classical result. Note, however, that the exact beta function
is proportional to $\p_{a_i}\p_{a_j}\langle\Tr\phi^2\rangle$, which includes
instanton corrections. The above discussion also applies to all the
other invariants $s_k$, and the absence of logarithms,
which would have appeared in perturbative contributions, is necessary
for them to be globally defined coordinates on the quantum moduli space.
\bigskip

Let us now turn to the remaining classical groups with
$N_f$ hypermultiplets in the defining
representation $\underline{N}_C$\matterall. Here the
Riemann surfaces are given by curves of the form \matterall
\eqn\sospcurve{
x y^2=W^2+F
}
where now for $x\to\infty$,  $W\sim x^r$ and $F\sim x^{N_f+\nu}$
where $\nu=4,3,0$ for $SO(2r),\,SO(2r+1)$ and $Sp(2r)$, respectively.
The meromorphic
differential $\lambda$ is
\eqn\sosplambda{
\lambda={1\over 2\pi i}{W F'-2 W' F\over y F}\,dx
}
with the asymptotic behavior at infinity
$\lambda\sim{1\over 2\pi i}(l({\rm adj}.)-N_f l(\underline{N}_C))
{dx\over\sqrt{x}}$ where $l({\rm adj.})=2(N_C-2),\, N_C+2$ and
$l(\underline{N}_C)=2,\,1$ for $SO(N_C)$ and $Sp(N_C)$, respectively.
The combination of the indices of the representations
appearing in the asymptotic expression of $\lambda$, is exactly the
one-loop coefficient $b_1$
of the beta-function for an $N=2$ supersymmetric
gauge theory with $N_f$ hypermultiplets in the defining representation.
Introducing the local uniformization variable $x=1/\xi^2$ one finds that
(we are again only considering the massless case here)
\eqn\lcoeff{
\lambda_{-2}=-{1\over\pi i}\bigl(l({\rm adj.})
-N_f l(\underline{N}_C)\bigr)\,.
}
Likewise one finds the asymptotic behavior of
$\omega_k=\partial_{s_k}\lambda$ as $\omega_k=(\omega^{(k)}_0
+\omega^{(k)}_1\xi+\dots)d\xi$ with
$\omega_0^{(k)}=-{1\over\pi i}\delta_{k,1}$. Note that in the notation
of ref.\matterall\ $s_1$ is the quadratic invariant:
$s_1={1\over2}\Tr\langle\phi^2\rangle$.
Inserting this into the Riemann relation
\Riemann\ we get
\eqn\othergroups{
{\p\over \p s_k}(\L\p_\L{\cal F})={2i\over\pi}\bigl(l({\rm adj.})
-N_f l(\underline{N}_C\bigr)\delta_{k,1}\bigr)
}
\medskip
Let us now turn to the massive case.
Here we have to use the Riemann bilinear relation for one abelian
differential of the first kind ($\omega_k$)
and the other of the third kind ($\lambda$) with
first and second order poles. We will concentrate on the case of
$SU(N_C)$. The other groups can be treated similarly.
In fact, the meromorphic differential $\lambda$ now has simple poles at
$x_i=m_i$ with residues $m_i$ and a double pole at infinity
where it behaves as
\eqn\lambdamassive{
\lambda=(\lambda_{-2} w^{-2}+\lambda_{-1} w^{-1}+\lambda_0+\dots)dw
}
with
$$
\lambda_{-2}={1\over 2\pi i}(2N_C-N_f) \qquad{\rm and}\qquad
\lambda_{-1}=-{1\over 2\pi i}\sum_{i=1}^{N_f}m_i
$$
The relevant bilinear relation gets contributions from both
of these coefficients as well as from the residues of the poles
at $x_i=m_i$. The contribution from $\lambda_{-2}$ is the same as
in the massless case. The contribution from the poles at $m_i$ and
the pole at infinity is
\eqn\residues{
2\pi i\sum_i{\rm res}_{x_i} \lambda\int_{x_0}^{x_i}\omega_k
+2\pi i\,{\rm res}_\infty\lambda
\int_{x_0}^\infty\omega_k=-\sum_{i=1}^{N_f} m_i\int_{m_i}^\infty\omega_k
}
where $x_0$ is an arbitrarily chosen point on the Riemann surface
\foot{The independence of the choice follows from the fact that the
residues
of meromorphic differentials on Riemann surfaces sum up to zero.
For more details on this relation, see refs.\rs.}.
This leads to
\eqn\result{
{\p\over \p s_k}\left(\L\p_\L+\sum_i m_i\p_{m_i}\right){\cal F}
={i\over\pi}(2 N_C-N_f)\delta_{k,2}-\sum_i m_i\,\int_{m_i}^\infty\omega_k
}
Recall that $\omega_k=\p_{s_k}\lambda$ so that this relation
can be integrated
w.r.t. $s_k$ leading to a generalization of eq.\dfdl:
\eqn\dfdlsun{
(\L\p_\L+\sum_i m_i\p_{m_i}){\cal F}
={i\over 2\pi}(2 N_C-N_f)\langle\Tr\phi^2\rangle-
\sum_i m_i\int_{m_i}^\infty\lambda\,.
}
Note that now, in contrast to the massless case,
the right hand side seems to depend on all the moduli
$s_k$.
We have not attempted to do the remaining integrals explicitly.
But let us demonstrate that this expression has in
fact the correct decoupling limit. We decouple one of the hypermultiplets
by taking the limits, say, $m_{N_f}\equiv M\to\infty,\,\L_{N_f}\to 0$
while
keeping $\L_{N_f-1}^{N_C-N_f+1}=M\L_{N_f}^{N_C-N_f}$ fixed.
To perform the integral $-M\int_{M}^\infty\omega_k$ we first
change variables $x=M\tilde x$ and then perform the
decoupling limit. In this limit $y(x)\to M^{N_C} x^{N_C}$ and the
integral becomes ${i\over\pi}M^{2-k}\int_1^\infty
{d\tilde x\over \tilde x^k} \to {i\over\pi}\delta_{k,2}$.
The integrals for $i=1,\dots, N_f-1$ only change in such a way that
$\omega_k$ turns into the holomorphic differential appropriate for the
curve with $N_f-1$ flavors. We thus find that on the right hand side of
eq.\result\ we get the change $(2N_C-N_f)\to(2N_C-(N_f-1))$.
The left hand side changes as
$\L_{N_f}\p_{\L_{N_f}}+\sum_{i=1}^{N_f}m_i\p_{m_i}\to
\L_{N_f-1}\p_{\L_{N_f-1}}+\sum_{i=1}^{N_f-1}m_i\p_{m_i}$.

\medskip

Let us now briefly mention that in all cases where $\L\p_\L{\cal F}$
is proportional to
$u$, $u$ is in fact invariant under $Sp(2r;{\bf Z})$ transformations
$\pmatrix{a_D\cr a\cr}\to\pmatrix{\tilde a_d\cr\tilde a\cr}
=\pmatrix{A&B\cr C&D\cr}\pmatrix{a_D\cr a\cr}$.
This is
essentially proven in \Matone\ for $SU(2)$.
A simplified version of his proof
can be easily generalized to arbitrary groups.
{}From $\tilde{\cal F}(\tilde a)
=\tilde{\cal F}(\tilde a(a))$
it follows that $\p_{a^j}\tilde{\cal F}(\tilde a(a))
=\left({\p\tilde a^i\over\p a^j}\right)\tilde a_{D_i}$.
This relation can be integrated
to yield
\eqn\modular{
\tilde{\cal F}(\tilde a)={\cal F}(a)+{1\over2}a^T B^T D a+
{1\over 2}a_D^T C^T A a_D
+a^T B^T C a_D
}
This implies that ${\cal F}-{1\over 2}a^T a_D={1\over2}\L\p_\L{\cal F}$
is invariant.

\medskip
Finally, one may consider $N=2$ models in their Coulomb phase also for
matter superfields in representations other than the adjoint or
fundamental representations.
For those cases it is plausible that the
$b_1$ factor in eq.\dfdlsun\ will be  replaced  by
$2N_c-\sum_i l_i({\rm matter})$.

The gauge kinetic terms of the low energy effective action  of
supersymmetric gauge theories   in their   Coulomb phase can be determined
from hyperelliptic curves not only for   $N=2$  supersymmetric
models but
also for  $N=1$ ones\IS.
As in the $N=2$ case, the ground state of these $N=1$ models
is described by  an hyperelliptic quantum  moduli
space characterized by its
singularities and monodromies.
The determination of the curve follows from
the classical singularities, instantons corrections and the global
symmetries
of the theory.
For instance  the curves which correspond to $SU(N_C)$
$N=1$ models with one adjoint
representation and $N_f$ fundamentals
(denoted by $(N_{ad}=1,N_f)$) takes the form of eq.\susurface \Hoz.
 The corresponding polynomial  $F$ is given now by
 $F=F(x,\Lambda, m_{ij},Y_{ij})$
where  $m_{ij}$ and $Y_{ij}$ are the  quark mass matrix and the matrix of
 Yukawa couplings.
 When  $Y$ is a unit matrix and  $m$ is diagonal
the model admits an  additional supesymmetry.  The curves in that case
turn into those  of $N=2$ models with $N_f$ hypermultiplets.

Starting  with a curve that corresponds to a given $N=1$ model
in its Coulomb phase  one can
follow the same steps taken above
and prove an analogous relation to the one given in eq.
\dfdlsun.  We now discuss the relation for certain $N=1$
classes of models.
 Using the curves of \Hoz, it turns out that  for the class of
 models $(1,N_f)$ the relation
is the same as that given in eq.\dfdlsun apart from a replacement of
$m_i$ by the
eigenvalues of the matrix $Y^{-1}m$.

In case of  $(2,0)$  $N=1$  models
the condition for a Coulomb phase is that the determinant of
the adjoint mass matrix vanishes\IS.
The curve for $SU(N_C=2)$\IS
 is identical to that of the $N=2$ case with $N_f=0$
when one replaces $\Lambda^2_{N=2}$ with $\half\Lambda_{N=1}m_{ad}$.
where $m_{ad}$ is the mass of the massive adjoint superfield.
A similar situation occurs in the $(2,1)$ model\EFGR.
 We  therefore anticipate that the $(2,N_f)$ curves will
coincide with those of the $(1,N_f)$ models by a substitution of
$\Lambda_{N_{ad}=1}^{2N_c-N_f}
\sim\tilde\Lambda_{N_{ad}=2}^{N_c-N_f}m_{ad}^{N_c}$

Naively, it seems that the l.h.s of the relation, for instance
for $N_f=0$, takes the form of
$\tilde\L\p_{\tilde\L}\F+ m_{ad}\p_{m_{ad}}\F$, and on the r.h.s
the term proportional to $s_2$
involves the $b_1$ pertaining to
 the one adjoint case.  This is quite surprising since apriori
 we expect such a $b_1$ to appear
only when the massive adjoint decouples.
The full determination of the relation and the decoupling
 for this class of models as well as  those  which involve
 other representations is under current
investigation.

\medskip
The relation discussed in this paper appears as a simple partial
differential equation
for the prepotential ${\cal F}$. In order to determine ${\cal F}$
completely one
needs more equations. Already in the pure $SU(2)$ case one needs
one more independent
relation. It would be great if one could obtain  enough relations
which would, in turn,
determine ${\cal F}$ in a simple way.

Finally we note that while the local counterpart of this relation
seems to be
quite important \KKLMV, the full physical meaning of the relation
still alludes us.
For fixed $\L$, in the massless case,  we can rewrite it as
\eqn\andim{
(\sum_i a_i \p_{a_i}-2){\cal F}={1\over 2\pi i} b_1\langle\Tr\phi^2\rangle
}
This equation looks completely quantum mechanical.
Moreover, as discussed in this
letter, its non-trivial content is associated with the non-perturbative
contributions on both sides. The left hand side of \andim\ looks
as if it is
related to the
``anomalous dimension"
of ${\cal F}$, i.e. to the deviation of
${\cal F}$ from its classical dimension 2. This is due to
quantum effects associated
with the appearance of $\L$. The right hand side involves
the beta function.
It is tempting to think that one could understand this relation
in terms of RG ideas.
So far we have not been successful in doing it.

\vskip2truecm

\bigskip
\noindent{\bf Acknowledgement:} One of us (S.T.) would like to thank
F. Ferrari and W. Lerche for valuable discussions and
Tel Aviv University for
hospitality. S.Y. would like to thank G. Veneziano for
interesting discussions.

\listrefs
\vfill\eject

\bye